\documentclass[12pt]{article}

\advance\hoffset by -1.1truecm
\advance\voffset by -1.0truecm

\begin{document}
\title{A Generalized Shannon Sampling Theorem, \\Fields at the
 Planck Scale as
Bandlimited Signals}
\author{Achim Kempf\\
Institute for Fundamental Theory, Department of Physics\\
University of Florida, Gainesville, FL 32611, USA\\
{\small Email:  kempf@phys.ufl.edu}}

\date{}

\maketitle

\vskip-8truecm

\hskip11.5truecm
{\tt UFIFT-HEP-99-04}

\hskip11.5truecm {\tt hep-th/9905114}
 \vskip7.1truecm

\begin{abstract}
It has been shown that space-time coordinates can exhibit only
very few types of short-distance structures, 
if described by linear operators: they can be continuous, discrete or
``unsharp" in one of two ways. 
In the literature, various quantum gravity 
models of space-time at short distances
point towards one of these two types of unsharpness.
Here, we investigate the properties of fields 
over such unsharp coordinates.
We find that these fields are continuous  - but possess
only a finite density of degrees of freedom, similar to fields on lattices.
We observe that this type of unsharpness 
is technically the same as 
the aperture induced unsharpness of optical
 images. It is also of the same type 
as the unsharpness of the time-resolution of
 bandlimited electronic signals.
Indeed, as a special case we recover
the Shannon sampling theorem of information theory.
\end{abstract}

At the heart of every candidate theory of 
quantum gravity is an attempt to
understand the structure of space-time at 
very short distances. The reason is a
simple gedanken experiment: 
the latest when trying to resolve distances 
as small as the Planck scale
the accompanying energy-momentum fluctuations 
due to the uncertainty relation
should cause curvature fluctuations large enough
 to significantly disturb the
very space-time distance which one attempts to resolve. 
Speculations about the resulting behavior of  
space-time at small distances have ranged from the idea that
space-time is discrete, to that it is foam-like, 
to that space-time may be a 
derived concept with a highly dynamical short-distance structure,
as e.g. string theory would suggest.
At least at present, 
however, there is no experimental access to
sufficiently small scales, and therefore, 
a priori, the short-distance structure
of space-time could still be any one out of infinitely many
possibilities.

In this context, it has recently been pointed out, 
in \cite{ak-erice}, that the range of possible 
short-distance structures can be reduced to only very few 
basic possibilities, under a certain assumption. The assumption is  
that the fundamental theory
of quantum gravity possesses for each 
dimension of space-time an operator $X^i$ 
which is linear and whose expectation values are real. 
The dynamics of these $X^i$ may be complicated 
and the $X^i$ may or may not commute.   
Nevertheless, one can prove on functional analytic grounds that
any such operator $X^i$, considered separately,
 describes a coordinate which is 
necessarily either continuous or discrete, or it
 is unsharp in one of two
well-defined ways. All other cases are mixtures of these.

Since continua and lattices are familiar, 
we will here study one of the two types of 
unsharp short-distance structures. 
The second type of unsharpness will be dealt with elsewhere.  
The type of unsharp coordinate which we will here investigate
can be characterized by an uncertainty relation \cite{ak-erice}:
Such a coordinate is described by an operator $X^i$ for which the
formal standard deviation 
$\Delta X^i = \langle(X^i-\langle X^i\rangle)^2 \rangle^{1/2}$ obeys
some positive lower bound:
$$ 
\Delta X^i(\phi) \ge \Delta X^i_{min}(\langle 
\phi \vert X^i\vert \phi \rangle)
$$ 
Here, $\phi$ is any vector on which the operator can act, and 
the function $\Delta X^i_{min}(x)$ describes
how the lower bound depends on the $X^i$- expectation value.
If this were nonrelativistic quantum mechanics, 
the interpretation would be
that the $X^i$- coordinate is unsharp in the sense that
particles cannot be localized to arbitrary precision
 on the $x^i$- axis and that the lower bound
on the position resolution 
depends in general on the $x^i$- expectation value, i.e. on 
where on the $x^i$- axis one tries to localize the particle. 
The function $\Delta X^i_{min}(x)$
may in general also take the value zero, but we
 will here focus on the case where 
it is strictly positive.

This type of unsharp short-distance structure has indeed frequently
appeared  in quantum gravity and in particular in string theory. 
For example, 
several studies, see e.g. \cite{ucrrefs}, suggest that the Heisenberg
uncertainty relation may effectively pick up Planck scale or
string scale correction terms of the form:
\begin{equation}
\Delta x \Delta p ~\ge \frac{\hbar}{2}\left(1+ 
\beta (\Delta p)^2 + ... \right)
\label{ucr}
\end{equation}
For $\beta$ positive, the lowest order correction in
Eq.\ref{ucr} implies that there is 
a constant lower bound for $\Delta x$, namely
$\Delta x_{min} = \hbar \sqrt{\beta}$. 
Of course, it is not necessarily surprising if even quite
different candidate quantum gravity theories
arrive in this way or another 
at some positive lower bound $\Delta X^i_{min}(x)$ 
on the formal uncertainty in coordinates $X^i$,
 because, as we mentioned,
for real entities which are described by linear 
operators this is one out very few possibilities.

Our aim here is to investigate 
what this general type of unsharp short-distance structure
means in field theory: Is it possible to
define fields $\phi(x^i,y)$ 
``over" such an unsharp coordinate $X^i$? 
The operator $X^i$ should act simply as 
$X^i: \phi(x^i,y) \rightarrow x^i \phi(x^i,y)$  
while we let $y$ stand collectively for all other 
coordinates (if commutative) or any other quantum numbers.
The main question is, 
how do the fields depend on $x^i$, given that
an unsharp coordinate $x^i$ is
neither continuous nor discrete?
How does one calculate the Hilbert space scalar product of fields - 
does it involve an integral over $x^i$, 
a sum over discrete points on the $x^i$-axis, or something else?

As we will show here, the answer is that fields $\phi(x^i,y)$ over  
such unsharp coordinates are indeed well-defined:
these fields are continuous functions $\phi(x^i,y)$
over a continuous variable $x^i$. Crucially, however, these fields are
automatically ultraviolet cut off in the sense that
they possess only finitely many degrees of freedom 
per unit length along the $x^i$ coordinate, 
similar to fields on lattices! 

Before we begin describing the details, let us agree to from now on 
suppress the index $i$ and the other variables $y$.
We should also
mention that some of the operators which describe
unsharp coordinates of this type can only be represented on fields which 
possess isospinor indices, but this phenomenon
 will be discussed elsewhere. 

Let us begin with two definitions: 
By a \it discretization \rm of the
$x$-axis we mean a discrete set  of real numbers, $\{x_n\}$,
where $x_{n+1}>x_n$ and where $n$ runs through all integers.
By a \it partitioning \rm of the $x$-axis we mean a smoothly parametrized
family of discretizations $\{x_n(\alpha)\}$
which together make up the entire $x$-axis, namely such that
every point on the $x$-axis, i.e. every real number,
 occurs in exactly one of the discretizations. 

Now our claim is that to each unsharp coordinate $X$, as
characterized by a curve 
$\Delta X_{min}(x)$,
there corresponds a 
partitioning $\{x_n(\alpha)\}$ of the $x$-axis such that if 
a field $\phi(x)$ is known only
on one of the partitioning's discretizations 
then the field can already be reconstructed 
everywhere on the $x$-axis. Namely, if for some arbitrary fixed $\alpha$ 
the amplitudes $\phi(x_n(\alpha))$ are known for all $n$
then $\phi(x)$ can be recovered for all $x$ 
through a reconstruction formula of the form:
\begin{equation}
\phi(x) = \sum_n G(x,x_n(\alpha))~\phi(x_n(\alpha)) 
\label{recon}
\end{equation}
Thus, the knowledge of a field's amplitudes at finitely many points
per unit length along the $x$-axis
indeed suffices to describe the field entirely. 
Thereby, the operation of reconstructing a field is
 interchangeable with the operation 
of multiplying it by $X$:
$$ 
x \phi(x) = \sum_n G(x,x_n(\alpha))~x_n(\alpha) ~\phi(x_n(\alpha)) 
$$ 
The scalar product of two fields (as far as the $x$- 
dependence is concerned) 
is a sum:
$$ 
\langle \phi_1\vert \phi_2\rangle
= \sum_n \phi_1^*(x_n(\alpha))\phi_2(x_n(\alpha))
$$ 
This scalar product formula gives in fact the same result 
independently of $\alpha$, i.e. 
independently of the choice of discretization
on which the sum is being calculated.

Similarly, also the $X$- expectation value and the second moment of fields
can be calculated on any one of the discretizations $\{x_n(\alpha)\}$
and the result does not depend on $\alpha$. 
Correspondingly,  
$\Delta X(\phi) = (\langle\phi \vert X^2\vert\phi\rangle -
\langle\phi\vert X\vert\phi\rangle^2)^{1/2}$ is the 
standard deviation of the fields' discrete 
samples on any one of the discretizations 
$\{x_n(\alpha)\}$ of the $x$-axis.
We remark that, more generally,  if a field is not only 
 in the domain of $X$ but also in the domain of 
higher powers of $X$, say $X^r$, i.e. if the field decays
 at infinity with the corresponding 
inverse power, then the higher moments up to the $2r$'th are finite,
and they too are independent of the discretization 
in which they are calculated:
$$ 
\langle \phi\vert X^r \vert\phi\rangle~=~
\sum_n ~(x_n(\alpha))^r ~\phi^*(x_n(\alpha))\phi(x_n(\alpha))
$$ 
We now still need to address the question exactly 
how the minimum position uncertainty curve $\Delta X_{min}(x)$
corresponds to a partitioning of the $x$- axis.
One expects of course that 
in regions of the $x$-axis where $\Delta X_{min}(x)$ is small 
the spacing needs to be tighter and vice versa.

To see the precise relationship, let us first recall
 the minimum position uncertainty curve for 
particles which live on a one-dimensional lattice $\{x_n\}$.
Clearly, these particles can be localized to 
absolute precision $\Delta X=0$ at each of the 
lattice sites, say $x_{n_0}$, namely with the 
wave-function $\phi(x_n)=\delta_{n,n_0}$.
If, however, a particle's expectation value lies
 in between two lattice sites then
its standard deviation cannot be lower than some
 finite value. As is straightforward
to verify, the curve $\Delta X_{min}(x)$ for a
 one-dimensional
lattice consists of half-circles which arc from 
lattice site to lattice site. 

The fields over an unsharp coordinate do not live
on only one discretization of the $x$-axis,
but simultaneously on a whole family of
discretizations which together constitute a partitioning of
the $x$-axis. In 
contrast to ordinary fields over a lattice, 
fields over unsharp coordinates 
therefore obey an equation of the form (for arbitrary fixed $\alpha$):
\begin{equation}
\sum_n f_n(\alpha) \phi(x_n(\alpha)) = 0
\label{bcabstr}
\end{equation}
Eq.\ref{bcabstr} expresses that on each
 one of the discretizations the fields 
cannot be too peaked: We will find that 
$f_n(\alpha)\neq 0$ for all $n$,
which implies, for example, that fields
 $\phi(x_n)=\delta_{n,n_0}$ do not
occur. More precisely,
Eq.\ref{bcabstr} implies that the variable 
lower bound $\Delta X_{min}(x)$ is 
the joint lower bound of all the minimum 
$X$-uncertainty curves of 
the individual discretizations in the partitioning. 
Namely, if we denote the minimum $X$-uncertainty
 curve of the discretization 
to the parameter 
$\alpha$ by $\Delta X_{min}(x,\alpha)$
 (composed of half-circles which
arc from point $x_n(\alpha)$ to point 
$x_{n+1}(\alpha)$ for all $n$) then:
$$ 
\Delta X_{min}(x) = \max_{\alpha}\Delta X_{min}(x,\alpha)
$$ 
In this way, every partitioning $\{x_n(\alpha)\}$ of the $x$-axis
determines a minimum position uncertainty curve 
$\Delta X_{min}(x)$ and vice versa. We can describe partitionings 
conveniently by how their lattice spacings vary over the $x$- axis.
Indeed, for each partitioning there is a unique lattice spacing 
function $s(x)$ which obeys for all $n$ and $\alpha$:
$$ 
s((x_{n+1}(\alpha)+x_n(\alpha))/2)
 = x_{n+1}(\alpha)-x_n(\alpha)
$$ 
Its inverse, $\sigma(x):= 1/s(x)$,
the ``density of degrees of freedom" function, of course also
describes an unsharp coordinate entirely.

Interestingly, $s(x), \sigma(x)$ and, correspondingly,  the
minimum position uncertainty curve $\Delta X_{min}(x)$ 
cannot vary arbitrarily abruptly. Intuitively, the reason is clear:
if a particle can be localized
only to very little precision around one point 
on the $x$-axis, then it is plausible that the  
particle cannot be localized to very high precision 
around a closely neighboring point. 

In fact, we find that 
the possible spatial variability of the unsharpness of a coordinate
is constrained to the extent that
one discretization, say $\{x_n(0)\}$, together with 
the set of data $\{\frac{d}{d\alpha}x_n(0)\}$, i.e. together with 
the discretization's  derivative with respect
to $\alpha$, already determines an entire partitioning $\{x_n(\alpha)\}$.
(Technically, the discrete amplitudes 
$v(x_n(0)):=(-1)^n (x_n(0) -i)^{-1} (dx_n/d\alpha(0))^{1/2}$
belong to a field $v(x)$ which can be 
reconstructed through Eq.\ref{recon},
thereby yielding $dx_n(\alpha)/d\alpha$ and therefore $\{x_n(\alpha)\}$ 
for all values of $\alpha$.)

Any unsharp coordinate can therefore be 
specified entirely by specifying 
one of its discretizations $\{x_n(0)\}$ together with its derivative 
$\{\frac{d}{d\alpha}x_n(0)\}$. Let us abbreviate these data as
$x_n := x_n(0)$ and $x^\prime_n: = 
dx_n(\alpha)/d\alpha \vert_{\alpha=0}$.

We still need to give explicit expressions for the 
coefficients $f_n(\alpha)$ of Eq.\ref{bcabstr}  
and of course also for the reconstruction 
kernel $G$ of Eq.\ref{recon}. 
Expressed in terms of the data $\{x_n\}$ and $\{x^\prime_n\}$,
we obtain (after lengthy calculation): 
\begin{equation}
f_n(0)~ = ~(-1)^n~ \sqrt{x_n^\prime}
\label{fn}
\end{equation}
and 
\begin{equation}
G(x,x_n) = (-1)^{z(x,x_n)} \frac{\sqrt{x^\prime_n}}{x-x_n}
\left(\sum_m \frac{x^\prime_m}{(x-x_m)^2}\right)^{-1/2}
\label{ker1}
\end{equation}
Here, $(-1)^{z(x,x_n)}$ provides a sign factor
such that $G(x,x_n)$ is continuous in $x$.
The sign factor arises naturally in a product representation:
$$ 
G(x,x_n) = \lim_{N\rightarrow\infty}
\frac{\prod_{\vert m\vert<N,m\neq n}(x-x_m)}{
\sqrt{\sum_{\vert r\vert<N} \frac{x_r^\prime}{x_n^\prime}
\prod_{\vert s\vert<N,s\neq r} (x-x_s)^2}}
$$ 
The proof of these results is rather
 technical. It is contained in a 
previous version, see \cite{ak-shannon-1}, 
and will be presented in detail in a 
follow-up paper. Let us here only sketch the proof:
The self-adjoint operator  $X(0)$ with
 purely discrete spectrum $\{x_n\}$
possesses simple symmetric restrictions $X$, each 
with a $U(1)$-family 
of self-adjoint extensions $X(\alpha)$. 
It can be shown that their spectra, $\{x_n(\alpha)\}$,
yield partitionings of the real line and that
the data $\{x_n^\prime\}$ suffice to 
specify the restriction and consequently the partitioning.
The main part of the proof then consists in calculating the 
unitaries which interpolate the eigenbases of the extensions. 
The matrix elements of those unitaries
 constitute the reconstruction kernel.

We eventually arrive at  
one-parameter resolutions of the Hilbert space identity 
in terms of an overcomplete and  continuously para\-metr\-ized set
of normalizable vectors: $$ 1=
\frac{1}{2\pi}\int_0^{2\pi}d\alpha\sum_n\vert
x_n(\alpha)\rangle\langle x_n(\alpha)\vert=\frac{1}{2\pi}~
\int_{-\infty}^{+\infty} dx~\frac{ d\alpha}{dx}~\vert
x\rangle\langle x\vert $$
Note that coherent states and continuous 
wavelets, see e.g. \cite{jrk}, yield analogous two-parameter 
resolutions of the identity. 

Let us now consider the instructive special case of unsharp 
coordinates whose minimum position uncertainty
curve $\Delta X_{min}(x)$ is  
constant. In this case,
also the density of degrees
of freedom $\sigma(x)$ is constant,  $\sigma
=(2\Delta X_{min})^{-1}$, and the
corresponding partitioning $\{x_n(\alpha)\}$ of the $x$-axis reads:
$$ 
x_n(\alpha) =  2 n \Delta X_{min} + \alpha 
$$ 
We read off that $x_n =x_n(0) = 2 n \Delta X_{min}$ 
and $x_n^\prime = \frac{d x_n}{d\alpha}(0) = 1$.
Applying these parameters in Eq.\ref{ker1} yields the 
reconstruction kernel. In this special case
here we can use the fact  that 
$$ 
\sum_{n}\frac{1}{(z-n)^2}=\left(\frac{\pi}{\sin \pi z}\right)^2
$$ 
to obtain a particularly simple expression for the kernel:
$$ 
G(x,x_n) = \mbox{sinc}\left(\frac{\pi (x-x_n)}{2\Delta X_{min}}\right) 
$$ 
We observe that the kernel, being a sinc-function, is the Fourier
transform of the function which is $1$ in the 
frequency interval $[-1/4\Delta X_{min},
+1/4\Delta X_{min}]$ and which vanishes everywhere else.  
This means that the set of fields over a 
coordinate with constant unsharpness
$\Delta X_{min}$ has a particularly simple 
characterization: It is the set of fields whose
frequency range is limited to the interval
 $[-\omega_{max},\omega_{max}]$, where
$\omega_{max} = 1/4\Delta X_{min}$. Also Eq.\ref{bcabstr} acquires
a simple interpretation: Eq.\ref{fn} yields $ f_n(0) = (-1)^n $ so that, 
as is readily verified, Eq.\ref{bcabstr}
 expresses that the fields' Fourier
transforms vanish at $\pm \omega_{max}$, i.e. 
Eq.\ref{bcabstr} is now a boundary condition
in Fourier space.

The fact that functions whose frequency range is within the
interval $[-\omega_{max},\omega_{max}]$
can be reconstructed everywhere, via the sinc-function kernel
$G(x,x_n)=\mbox{sinc}(2\pi(x-x_n) \omega_{max})$, from
their values on discrete points $\{x_n\}$ 
with spacing $1/2\omega_{max}$,
is indeed well-known, 
namely as the Shannon sampling theorem. 
The sampling spacing
$x_{n+1}-x_n=1/2\omega_{max}$ is called the Nyquist sampling rate.
The basic idea of the theorem was actually 
already known to Borel (1897) and,
according to \cite{marks}, perhaps even to Cauchy (1841). 

Shannon is credited for introducing the theorem
into information theory in the 1940s,
see \cite{shannon}: 
Shannon showed that, due to 
noise and other limitations, in effect
only finitely many amplitude levels of electronic signals 
can be resolved, say $N$.
Consequently, for any given ensemble of signals, 
the measurement of a signal's
amplitude at some fixed time $t$ can yield at most $\log_2N$ bits
of information. Crucially now, Shannon's ansatz is to idealize 
electronic signals $\phi(t)$ as \it bandlimited, \rm 
i.e. as frequen\-cy\--limited functions. The sampling theorem then
shows that $2\omega_{max}$ amplitude
measurements per unit time suffice to capture such signals entirely -
and this implies that these signals can
carry information at most at the rate $b= 2 \omega_{max}
\log_2 N$ in bits/sec or, in 
terms of the density of degrees of
freedom: $b=\sigma \log_2N$. 

The ability provided by the sampling theorem 
to reconstruct continuous signals from discrete samples
and the analysis of their information content
have indeed proven very
useful in ubiquitous applications from 
scientific data taking and data analysis to
digital audio and video engineering. 
This of course motivated several generalizations
 of the sampling theorem,
see e.g. \cite{jerrietc}. For example, there are methods to improve
the convergence of the reconstruction 
through oversampling, see e.g. \cite{marks}. 

One may ask, therefore, why it should have been
 difficult to generalize the theorem 
for time-varying information densities.
The main reason is that what would seem to be the obvious approach, 
namely to try to use Fourier theory to define a notion of
time-varying bandwidth, $\omega_{max}(t)$,
 faces major difficulties: Firstly, 
the resolution of a signal's frequency content
 in time is of course limited by
the time-frequency uncertainty relation. Secondly,  
even low bandwidth signals can actually
 oscillate arbitrarily fast in any interval of
finite size (on these so-called
 superoscillations, see e.g.  \cite{ahaakso}). 

We here avoid those problems by not even trying to define
variable bandwidths $\omega_{max}(t)$ in any Fourier sense.
Instead, we obtain a handle on variable information densities through
variable densities of degrees of freedom $\sigma(t)$, which are
well-defined directly in the time-domain. 
Possible practical applications are currently being explored.

We note that, as a by-product of considering the
 special case of constant density
of degrees of freedom we have found that 
the unsharpness of space-time according to the quantum gravity 
and string theory motivated 
uncertainty relation, Eq.\ref{ucr}, is indeed 
of the same type as the unsharpness
in the time-resolution of bandlimited electronic signals. 
In fact, it is also the same type as the fundamental
unsharpness of optical images since, as is well-known, 
the aperture induces a bandlimit on the measurement of angles. 
Of course, to find this type of unsharpness in such different contexts 
is again not necessarily surprising, given that unsharp real entities 
described by linear operators - within any arbitrary theory -
can exhibit only two types of unsharpness.

Our finding that fields over unsharp coordinates 
possess finite densities of degrees of freedom can serve, as we saw,
as the starting point for an information theoretic analysis of ensembles
of fields. This should be interesting to pursue. Indeed,
in studies in quantum gravity and in particular in string theory
the counting of degrees of freedom and 
an information theoretical perspective 
have recently found renewed interest, in particular in the contexts of
the black hole information loss problem
 and the holographic principle, see
e.g. \cite{big}.

Our observation that 
fields over unsharp coordinates are continuous but
behave in many ways like fields over lattices also raises 
questions such as, how do anomalies
 manifest themselves with this type of
ultraviolet cut-off: perhaps through fermion
 doubling as on lattices, or else?
Eventually, it should be possible to work out model independent 
phenomenological signatures of
this type of unsharp space-time. These might be testable if, as recent
models of large extra dimensions suggest possible, 
the onset of strong gravity effects is
not too far above the currently experimentally accessible
scale of about $10^{-18}m$, rather than
 at the Planck scale of $10^{-35}m$,
see e.g.\cite{arkanietc}.
\smallskip\newline
\bf Acknowledgement: \rm The author is grateful to John Klauder 
for very valuable criticisms.

\end{document}